\documentclass[10pt,halfline,a4paper]{ouparticle}

\usepackage{url}
\usepackage{palatino}
\usepackage{hyperref,cleveref}
\usepackage{amsthm}

\usepackage{xcolor}

\theoremstyle{definition}
\newtheorem{definition}{Definition}[section]

\newcommand{\scaption}[1]{\caption{\small #1}}

\begin{document}

\title{Sigmoids behaving badly: why they usually cannot predict the future as well as they seem to promise}

\author{%
\name{Anders Sandberg}
\address{Future of Humanity Institute, University of Oxford}
\email{\url{anders.sandberg@philosophy.ox.ac.uk}}
\and
\name{Stuart Armstrong}
\address{Future of Humanity Institute, University of Oxford}
\email{\url{stuart.armstrong@philosophy.ox.ac.uk}}
\and
\name{Rebecca Gorman}
\address{Berkeley Existential Risk Initiative}
\email{\url{rebecca.gorman@protonmail.com}}
\and
\name{Rei England}
\address{The Business School (formerly Cass), City, University of London}
\email{\url{solitude_saiyajin@hotmail.com}}
}

\abstract{Sigmoids (AKA s-curves or logistic curves) are commonly used in a diverse spectrum of disciplines as models for time-varying phenomena showing initial acceleration followed by slowing: technology diffusion, cumulative cases of an epidemic, population growth towards a carrying capacity, etc. Existing work demonstrates that retrospective fit of data is often impressive. We show that in time series data, the future fit tends to be poor unless the data covers the entire range from before to after the inflection point. We discuss the theoretical reasons for this: the growth data provides little information about the damping term (and vice-versa). As a consequence, forecasting with sigmoids tends to be very unreliable. We suggest some practical approaches to improving the viability of forecasting sigmoid models.}

\date{\today}


\maketitle

\begin{quotation}\noindent
\small A trend is a trend is a trend, \\
But the question is, will it bend? \\
Will it alter its course\\
Through some unforeseen force \\
And come to a premature end?

--Alexander Cairncross,  quoted in {\em New Statesman}, November 29, 1974.
\end{quotation}

\section{Introduction}


\begin{figure}
\begin{center}
\includegraphics[width=0.7\textwidth]{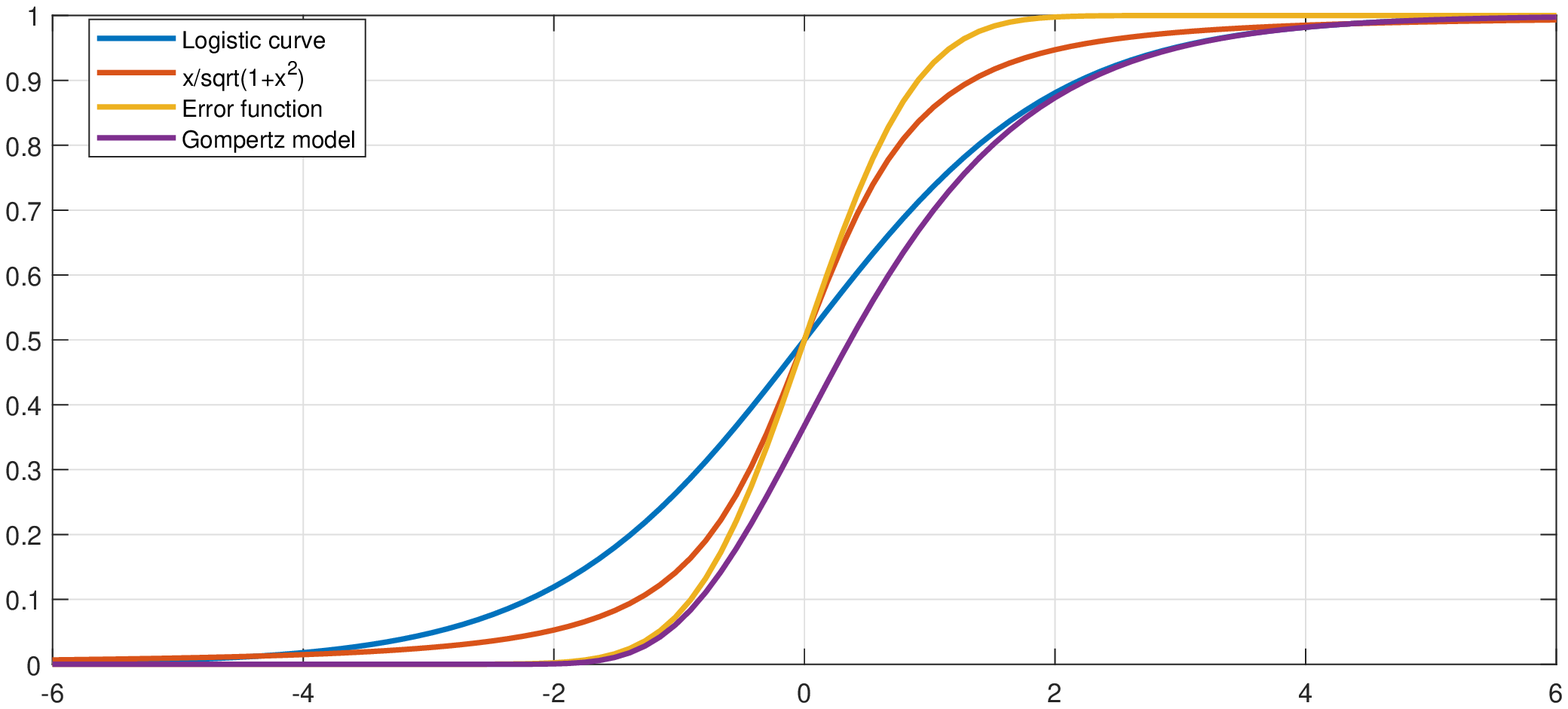}
\end{center}
\scaption{ Example sigmoid curves. The key properties are the time of the inflection point where it shifts from accelerating growth to slowing growth, the eventual asymptotic level, and the sharpness of the transition. Logistic curve $y(t)=L/(1+\alpha\exp(-\beta t))$, algebraic curve $y(t)=(L/2)(1+t/\sqrt{1+t^2})$, error function $y(t)=(L/2)(1+\text{erf}(\alpha t))$, Gompertz curve $y(t)=L \exp(-\alpha \exp(-\beta t))$ for $L=\alpha=\beta=1$ (actual curve models would also include a translation parameter $t_0$ setting when the inflection point occurs).
}
\label{fig:sigmoids}
\end{figure}

Sigmoids are popular models for many time-varying phenomena. They show a characteristic S-curve behaviour, with an initially accelerating growth period, leading up to an inflection point (or turning point/midpoint). After the inflection point, the growth slows and the curve tends towards an asymptote (or maximum/saturation level).

Many different phenomena can be modelled with sigmoids: technology diffusion, cumulative cases of an epidemic, spread of information, population growth towards a carrying capacity, etc (\autoref{fig:sigmoids}). There is a sizeable literature on their origins and how to fit data to them \cite{panik2014growth}. The retrospective fit of data is often impressive.


They are also often used for trend forecasting \cite{goldenberg2019s}. They are not just common motifs in time series, but  often have an underlying logic that is well understood (positive feedback that is reduced by a limiting factor, imitation between differently sized groups of technology adopters, saturation, etc.) and fit many domains well. 

In addition, mathematically they are easy to fit to data using modern software since they typically only have three parameters\footnote{Some functions are asymmetric, with a different curvature in the accelerating and slowing parts: this can add a fourth parameter. In many situations the asymptote ("100\%") can be known, removing a parameter.}. There are many functional forms and models giving similar shapes: --- logistic curves, error functions, Gompertz models, Bass curves \cite{bass1969new}, etc\ldots See \autoref{fig:sigmoids} for some examples. Paper \cite{young1993technological} discusses which functional forms are best for modelling given data.

This should be ideal for making principled, robust forecasts. Thus it is indeed a widely used method for time series forecasting in many domains, from technology diffusion \cite{harvey1984time,meade1998technological,massiani2015choice} over demographics \cite{raeside1988use} to sales forecasts \cite{bass1969new} and project management \cite{san2017s}. It often fits well with forecasting principles like "Structure the problem to use the forecaster's domain knowledge" and "Use a simple representation of trend unless there is strong contradictory evidence." \cite{armstrong2001extrapolation,meade2001forecasting}.

But others warned against this seeming rosy picture. In 1998, Nigel Meade and Towhidul Islam wrote:
\begin{quotation} "The straightforward policy of identifying an appropriate model and then using it to generate forecasts was shown to be difficult, if not impossible, to put into practice. The reason is that there  is insufficient evidence available in the time series data to permit model recognition. Data sets with known data-generating processes were simulated and used to demonstrate that measures of fit alone give inadequate guidance for model identification." \cite{meade1998technological} \end{quotation}
These warnings should have been heeded. There are very many cases where sigmoids have produced unreliable forecasts.

\begin{figure}
\begin{center}
\includegraphics[width=0.5\textwidth]{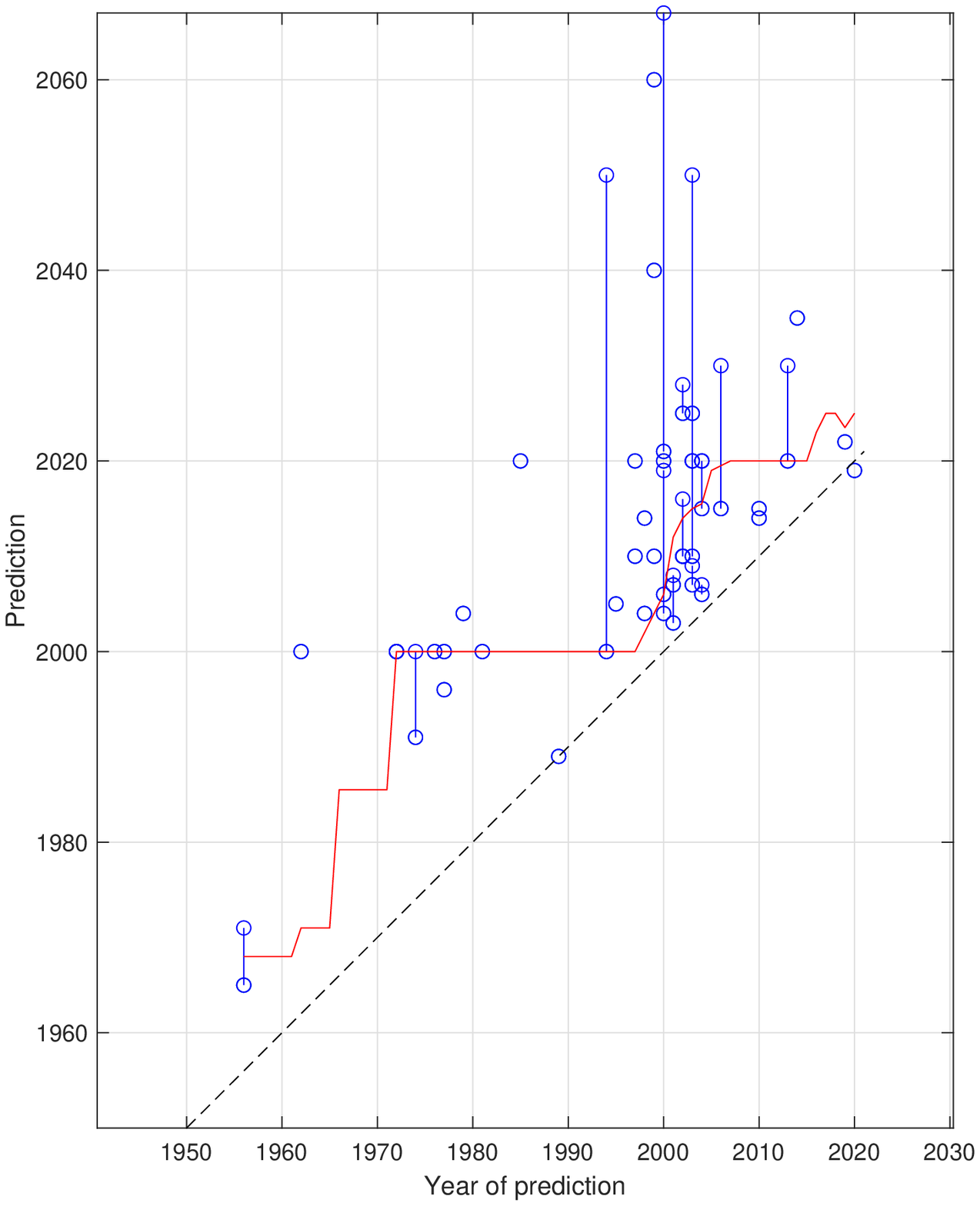}
\end{center}
\scaption{Predictions of peak oil production from different sources (compiled from Wikipedia:Peak Oil) 
showing year of the prediction and the predicted year when global oil production would start declining. Red curve shows median prediction at different years, based on forecasts not yet disproven. Not all predictions were achieved by using direct sigmoid fits. }
\label{fig:peakoil}
\end{figure}

A perennial example has been past "peak oil" predictions based on fitting the historical growth rate of oil production to a bell-shaped Hubbert peak, empirically often observed in actual regional oil extraction \cite{brandt2007testing}.\footnote{E.g. working with the derivative of the assumed sigmoid of cumulative oil extracted. In practice it is nearly always better to work with the cumulative curve for numerical stability reasons, which is what many fits do.} Hubbert in 1956 famously predicted peak US oil production around 1970 and global peak 1985-2000 \cite{hubbert1956nuclear}. Since then there has been a steady stream of papers and claims predicting an imminent peak, while actual global oil production has been increasing (see \autoref{fig:peakoil}).
These predictions take place in a fairly data-rich environment under the reasonable domain assumption that there is a finite amount to be extracted and that eventually cost will become prohibitive, and yet they have fairly consistently failed so far. 

While we do not have a comprehensive set of COVID-19 predictions there appears to be a plethora of failed predictions due to sigmoid models. Retrospective sigmoid fits to Chinese outbreaks worked well \cite{shen2020logistic}.
A data scientist fitted US COVID-19 cases during April 2020. As honestly recounted on his blog, at first it produced an excellent fit... and then not \cite{fotache2020predicting}. A peer-reviewed paper using more sophisticated methods published in July 2020 predicted a peak in late October 2020 and a cumulative  14.12 million infections worldwide \cite{wang2020prediction}, which sadly did not come true; the actual number of cases are already at least 10 times as high and the pandemic has lasted much longer. Even early papers taking known fitting problems explicitly into account produced an overly rosy picture of the future course of the pandemic (e.g. \cite{tatrai2020covid,lee2020estimation}). While some blame can be laid on governance failures and unmodelled second and third waves it is clear that predictions tended to consistently underestimate the global and local outbreaks and how early they would end. 

These examples show that for important forecasting problems sigmoids are being used in ways that consistently produce biased or misleading predictions. Typically, they (1) predict the inflection point where the curve starts slowing down to happen close in time to the present, and (2) are wildly uncertain about the eventual asymptote level.

A key insight should be: We cannot forecast a system with the same confidence with which we can backcast the same system. While backcasting, fitting a curve to a given dataset, can be very accurate and principled, forecasting means fitting a curve to a partial dataset where a non-random part (the future) is missing. This can introduce serious bias. Sigmoid curves are particularly sensitive to this problem despite their other good properties. 

This issue has been recognized by some people \cite{modis2007strengths,sandberg2014sigmoid,crozier2020forecasting,kucharavy2011application} but appear to have mainly been data-science "folklore" (or, "too obvious to mention"). In various application fields it has also been noted here and there (e.g. \cite{hsieh2004sars,shen2020logistic} note the fit to epidemiological data only becomes reliable after the time of peak case increase), but the generality of the problem has not been pointed out loudly enough. The purpose of this paper is to make it more widely known and look at some solutions. 

\section{The problems with predicting sigmoids}

Intuitively, the difficulty in predicting sigmoid comes from one simple fact: the initial data provides a lot of information about the initial growth rate, but very little about the asymptote and the inflection point.

This is difficulty is unavoidable due to the essential characteristics of a sigmoid: a function that has a growth term that causes increasing growth, followed by a transition to a damping term where growth is diminishing. The initial data will provide much information about the growth term, but very little (or none) about the damping term. This becomes a problem in two ways:

\begin{enumerate}
    \item Small amounts of noise in the initial data create great uncertainty about the asymptote and the inflection point.
    \item Even without noise, there are plausible families of sigmoid curves where the initial data might tell nothing about the asymptote and inflection point.
\end{enumerate}

\subsection{Sigmoids in general}

A sigmoid can be defined as a curve that starts close to $0$, grows faster and faster with time, until it hits an inflection point $t_0$ and starts growing slower and slower, eventually either reaching a limit value $L$, or asymptoting toward $L$. Typically this occurs due to a self-reinforcing growth process that weakens beyond a certain point. For a more formal definition, see \Cref{sigmoid:def} in \autoref{sigmoid:def:sec}.

To analyse this in more detail, let's look at the most common sigmoid example: the logistic curve.

\subsubsection{The logistic curve}
The logistic curve is a canonical version of the sigmoid. It is a solution to a differential equation of the form
\begin{eqnarray}
y'(t) = k\left(1-\frac{y(t)}{L}\right)y(t),
\end{eqnarray}
for positive constants $k$ and $L$. If we set $H(y)=1-y/L$, then this equation becomes
\begin{eqnarray}\label{H:eq}
y'(t) = k \cdot H(y(t)) \cdot y(t).
\end{eqnarray}

When $y(t)$ is small compared with $L$, then $H(y(t)) \approx 1$ and the differential equation is approximately
\begin{eqnarray}\label{expon:eq}
y'(t) \approx k y(t).
\end{eqnarray}

The solutions to \autoref{expon:eq} are pure exponential curves of equation $y(t)=C e^{kt}$ for some constant $C$. This is why the logistic curve starts off with a period of exponential growth, with growth rate $k$.

The logistic curve is defined by three parameters --- $k$, the growth rate, $L$, the asymptote, and $t_0$, the location of the inflection point. Knowledge of $t_0$ can be replaced by knowledge of $y(0)$, the value of the curve at $t=0$.

\subsubsection{General damping function}

As $y(t)$ grows, $H(y(t))=1-y(t)/L$ declines towards zero. This is the damping term in \autoref{H:eq}; it dominates after the inflection point, and is what limits $y(t)$ to asymptote to $L$.

A general damping term would be any decreasing $H$ such that $H(0)=1$ and  $H(y(t))$ tends to $0$ as $y(t)\to L$. We'll be a bit more specific: require that $H(y)>0$ for $y<L$, and the function $H(y)y$ (which, by definition, is $0$ at $y=0$ and $y=L$, and strictly positive between the two) has but a single local maximum for $y\in[0,L]$. Let $0 < y_0 < L$ be the $y$-value of this local maximum.

Then the differential equation
\begin{eqnarray}
y'(t) = k \cdot H(y(t)) \cdot y(t)
\end{eqnarray}
will define a sigmoid $y(t)$, with growth rate $k$, inflection point $t_0$ defined by $y(t_0)=y_0$, and asymptote $L$. Since it is a first order differential equation, it has a one-dimensional family of solutions; these solutions are just translated versions of each other, and such differ only in the value of $t_0$.

There is a converse to this result: any sigmoid in the sense of \Cref{sigmoid:def}, can be defined by such an $H$ and such a differential equation.  \autoref{diff:appendix} demonstrates these results.

\subsection{Unidentifiability of sigmoids from early data}

Thus a sigmoid can be characterised by the damping function $H$, the growth rate $k>0$, and, to fix the translation, by the value of $y$ at $t=0$.
A model would consist of a class of possible values for $(H,k,y(0))$. When modelling the process as a logistic curve, for instance, $k$ would be any strictly positive real, $y(0)$ would be any real number in $(0,L)$, and $H$ would be the family of function $1-y/L$, parameterised by $L$. Thus identifying a logistic curve from data involves identifying three parameters.

The logistic curve is real-analytic (its Taylor series converges and is equal to the function). Thus its global behaviour is determined by its local behaviour. So, in practice, the three parameters of the logistic curve can be determined by giving three noise-free values\footnote{
Technically, what matters is not that the logistic curve itself is real-analytic, but that for any $t_1$, $t_2$, and $t_3$, the values $y(t_1)$, $y(t_2)$, and $y(t_3)$ are real-analytic functions of the parameters $t_0$, $k$, and $L$.
} of $y(t)$.

But there is no reason to suppose that the underlying process is logistic. The logistic damping term models the damping as a pure linear function of the remaining population. But that need not be the case. For instance, if there are multiple populations with different reactions in a diffusion model, or if the network connecting members of the population is more complicated, then $H$ can take a very different form.

Even if we restrict to real-analytic families of $H$'s, there can be far more than three parameters to fit, and generically, all parameters would be relevant to estimating $L$.
But if $H$ is not real-analytic, it may be {\em impossible} to estimate $H$ from the initial data. 

\begin{figure}
    \centering
    \includegraphics[width=0.7\textwidth]{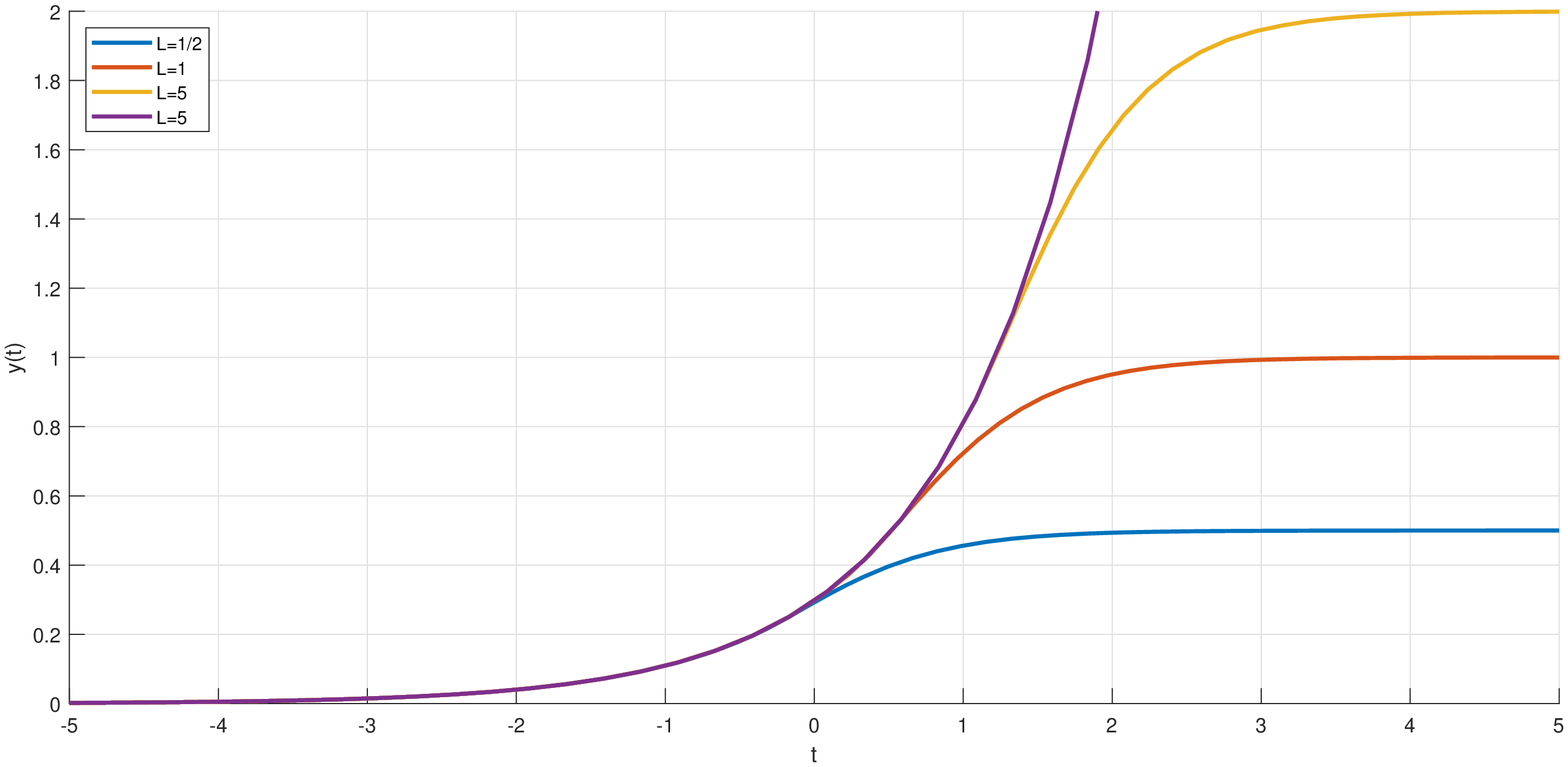}
    \scaption{Example of unidentifiability of sigmoids from early data: plots of sigmoids solving $y'(t)=\min(1,2-2y(t)/L)y(t)$, for $L=0.5$, $1$, $2$, and $5$.}\label{diff:L}
\end{figure}

The simplest none-real-analytic functions are piecewise functions. They can be regarded as crude approximations to many actual cases. If $H(y)=\min(1,2-2y/L)$, then $H$ is uniformly $1$ for $y < L/2$ --- and hence $y$ is a pure exponential until that value.
So if we considered the family of functions $H(y)=\min(1,2-2y/L)$, then different values of $L$ are completely indistinguishable for any data with $y<L/2$ --- even infinitely many noise-free samples will not distinguish them. The same goes for $y_0$, the $y$-value of the inflection point (which must be at least $L/2$), and hence the inflection point $t_0$ itself, which is defined by $y(t_0)=y_0$. This is illustrated by the curves in \autoref{diff:L} for $L \in \{0.5,1,2,5\}$; the curves match perfectly until they get close to their asymptote.

\begin{figure}
    \centering
    \includegraphics[width=0.4\textwidth]{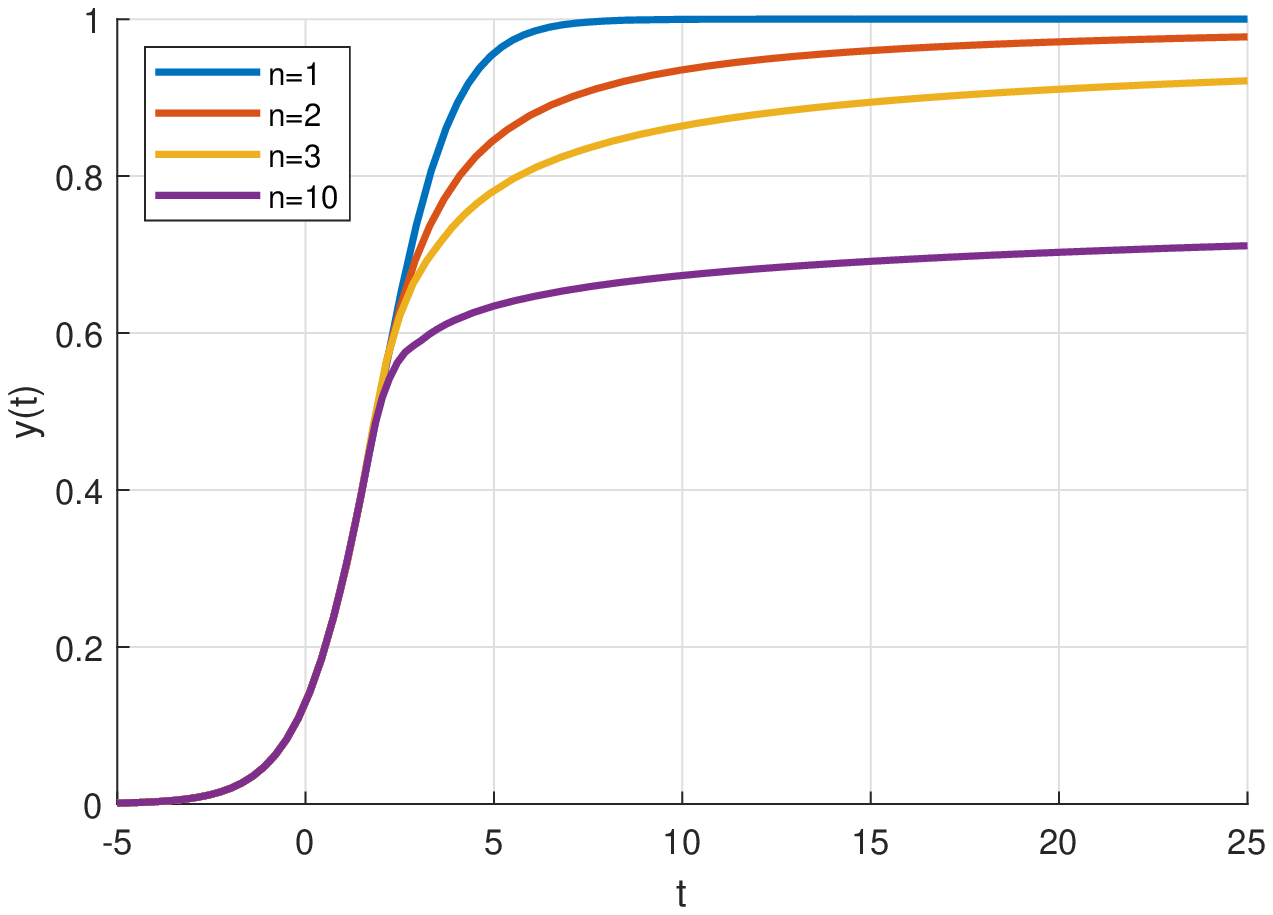}
    \includegraphics[width=0.4\textwidth]{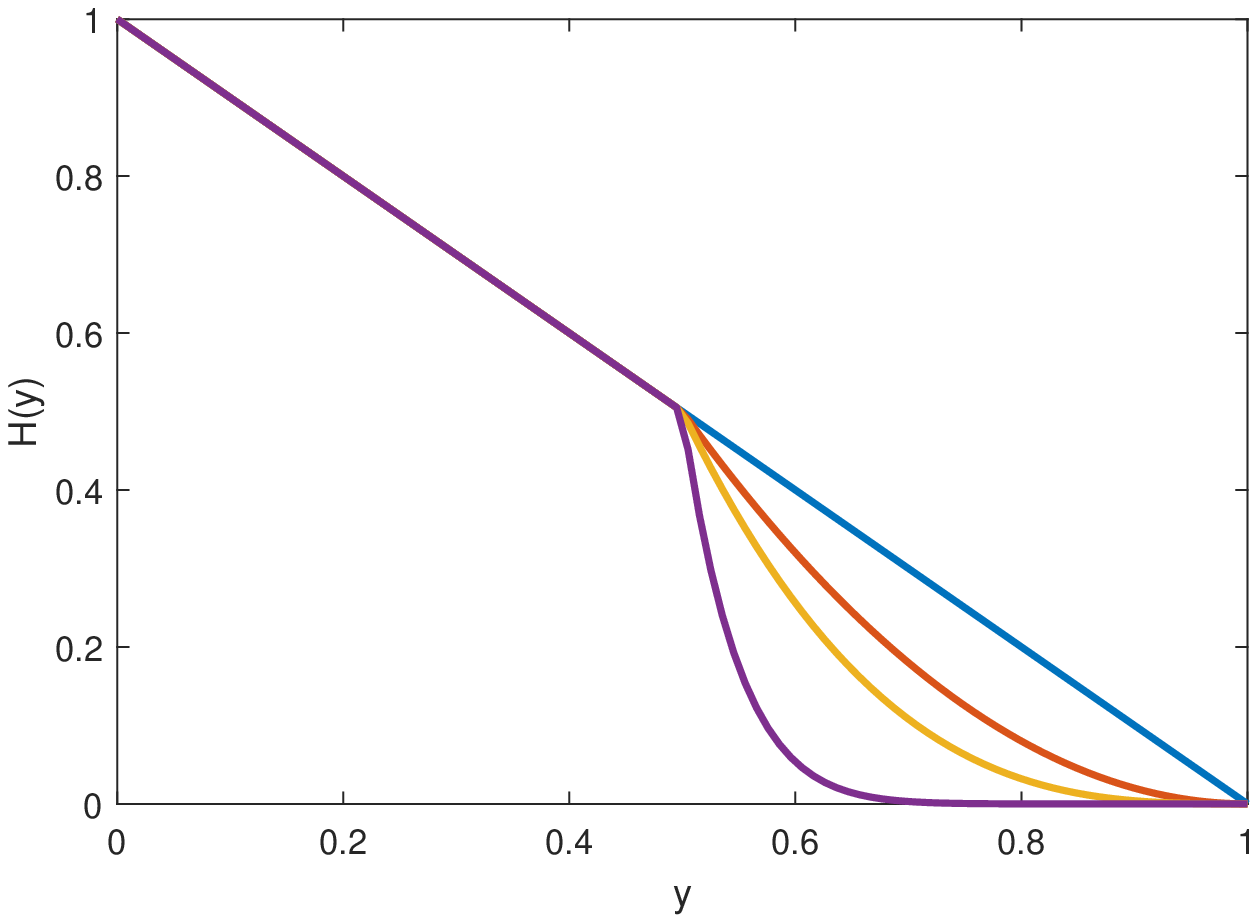}
    \qquad
    \includegraphics[width=0.4\textwidth]{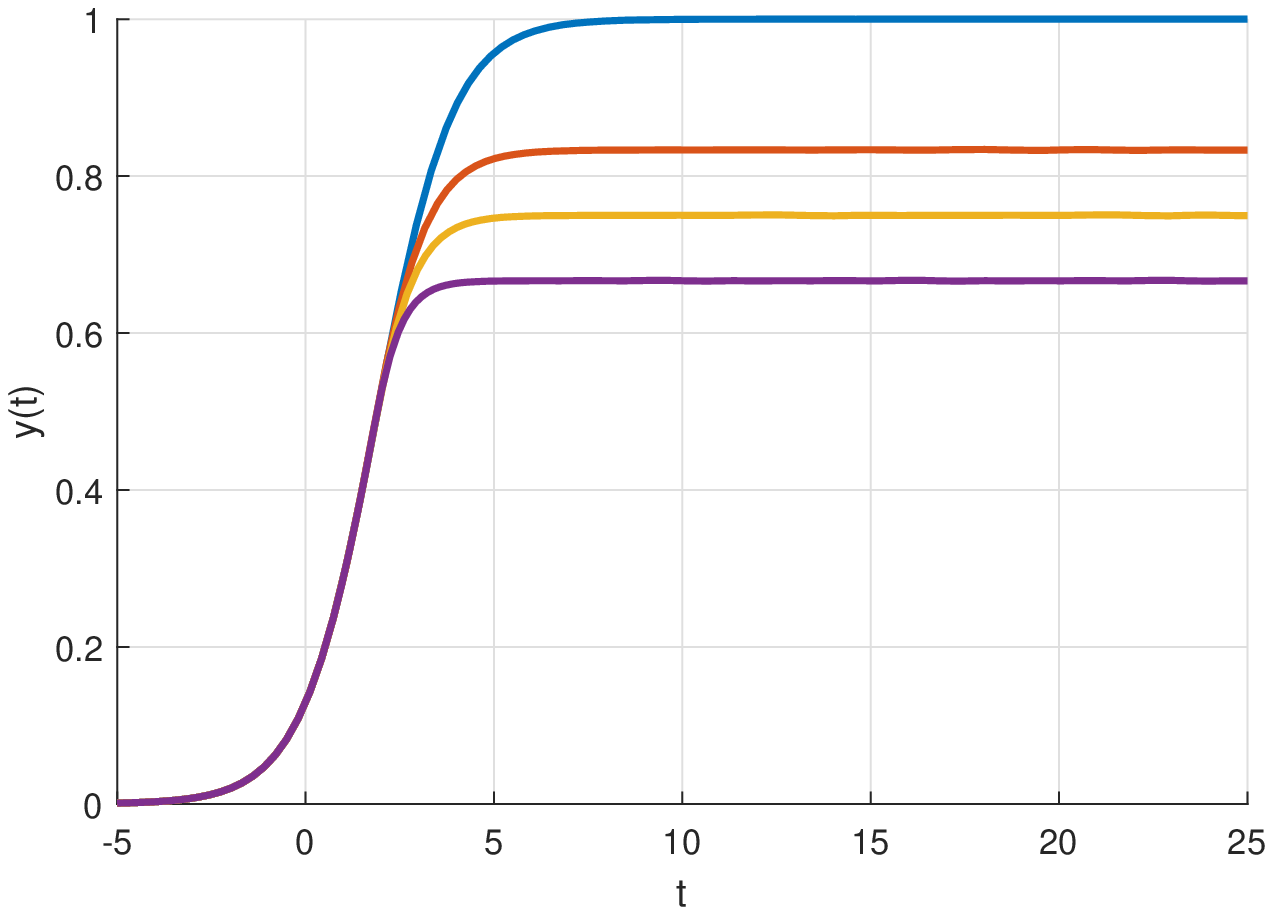}
    \includegraphics[width=0.4\textwidth]{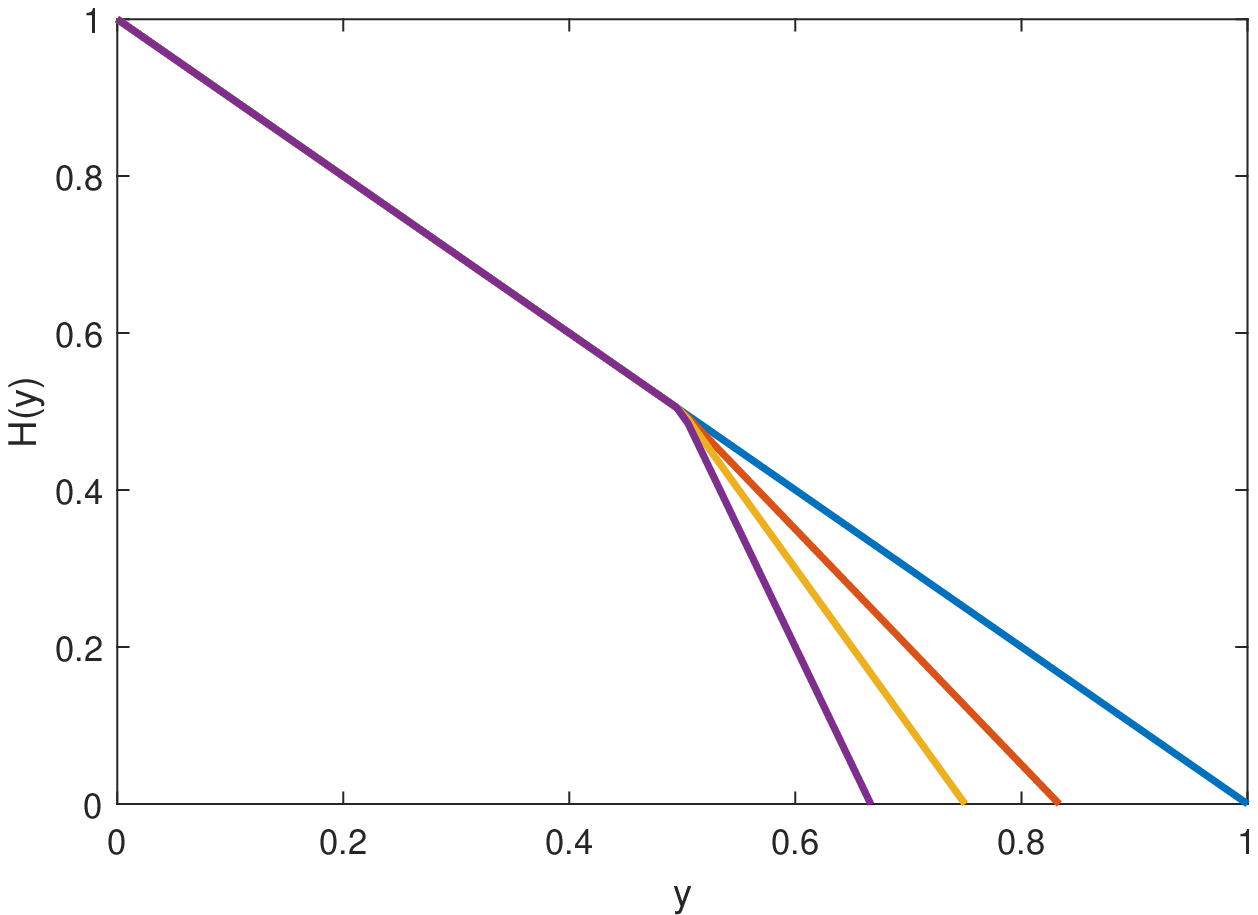}
    \qquad
    \caption{(Top) Curves with the same inflection points ($t_0=1.9$, $y_0=0.5$), same behaviour up to the inflection points, and same asymptote ($L=1$) but very different convergence rates to the asymptote. This is due to having different $H$s after $y(t)=1/2$, namely linear, quadratic, cubic, and tenth power convergence to $H(1)=0$.
(Bottom) Curves with the same inflection points ($t_0=1.9$, $y_0=0.5$), same behaviour up to the inflection points, but different  asymptotes. This is due to having different slopes of $H$s after $y(t)=1/2$ (slope $-1$, $-1.5$, $-2$, and $-3$).}
    \label{diff:t0}
    \label{diff:L:t0}
\end{figure}

\Cref{diff:t0} presents curves with the same asymptote ($L=1$), and the same behaviour up to their inflection point, but very different rates of convergence; this is due to $H$'s that start off linear but have a very different behaviour after $y(t)>1/2$. See figure~\ref{fig:unidentifiable} for an example of the same phenomenon in a contagion model where $H$ is emergent. Similarly, the lower plots in \Cref{diff:L:t0} present curves with same behaviour up to their inflection point, but different asymptotes.

A model is identifiable  if it is in principle possible to learn the values of the model's underlying parameters after obtaining an infinite number of observations from it (alternatively: all different parameter settings produce different probability distributions of data). It is unidentifiable if different parameter settings are observationally equivalent (e.g. they can generate the same distribution) \cite{rothenberg1971identification}.
This section has shown that general families of sigmoids are unidentifiable from early data.

\subsection{Weak identifiability of noisy sigmoids from early data}\label{weak:identifiability}

Real-analytic families with $n$ parameters means that the sigmoid can be identified from $n$ perfect noise-free samples. But real-analyticity doesn't tell us whether the sigmoid can be identified well from noisy data.

The challenge is the same as in the noise-free case: identifying the asymptote $L$ and the $y$-value of the inflection point, $y_0$, relies on identifying features of $H$ --- respectively, on identifying the point at which $H$ becomes $0$ and identifying the maximum of $H(y)y$.
These are typically `late' features of $H$: features beyond the scope of early data. So the question is whether these late features can be estimated from early data.

If $H$ comes from a family like $H(y)=\min(1,2-2y/L)$, the answer is a clear no. In that family, the late features are not even identifiable from noise-free early data; adding noise will only make the situation worse.

Even if the family is identifiable in principle, there are two challenges that need to be overcome:
\begin{enumerate}
    \item The early $H$ may not be very identifiable from early data.
    \item The late $H$ may not be very identifiable from early $H$

\end{enumerate}
Now, the identifiability of late $H$ from early $H$ depends on the choice of models (see \autoref{modelling}). As For identifying early $H$, note that \autoref{H:eq} implies that $H(y(t)) = y'(t)/(k(y(t))$. The $k$ can often be well estimated from the early data, but noise will interfere with the estimates of $y(t)$, $y'(t)$, and their ratio. Additive noise (where the sampled $\tilde{y}(t)$ is $y(t)+\epsilon$ for noise $\epsilon$) is more likely to cause errors in this ratio than multiplicative noise (where the sample $\tilde{y}(t)$ is $y(t)(1+\epsilon)$ for noise $\epsilon$). Errors are also to be expected if the noise distribution is misidentified; noise is generally modelled as Gaussian, but if the tails are fatter, Gaussian models can fail dramatically\footnote{Numerous cases of major loss in finance have been attributed to using Gaussian models when reality is heavy-tailed.

Linear least-square regression fails when errors are far from Gaussian due to the failure of the Gauss-Markov theorem to apply. Even when the noise model is of the appropriate type identifiability can still be weak. While it may take 30 observations in the Gaussian case to stabilize the mean up to a given level, it takes $10^{11}$ observations in the Pareto distribution case to bring the sample error down to the same level \cite{taleb2019probability}.}.

\begin{figure}
    \centering
    \includegraphics[width=0.7\textwidth]{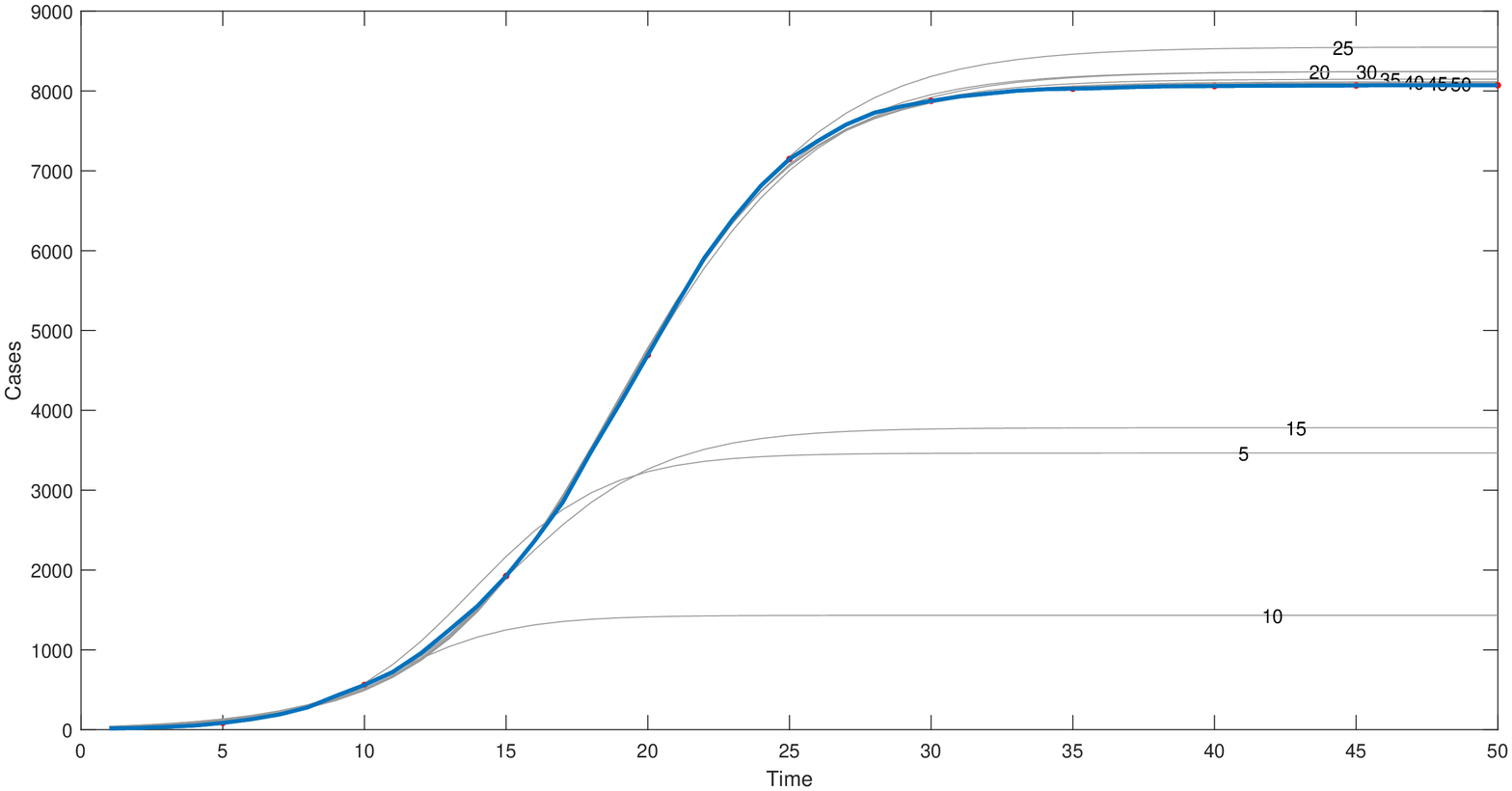}
    \qquad
    \caption{Example of instability of fit to the early part of data: cumulative case numbers from an agent-based model of contagion (blue) with fitted logistic curves (grey), each marked by when the forecast was made.}\label{fig:unidentifiable:1}
    \qquad
    \includegraphics[width=0.7\textwidth]{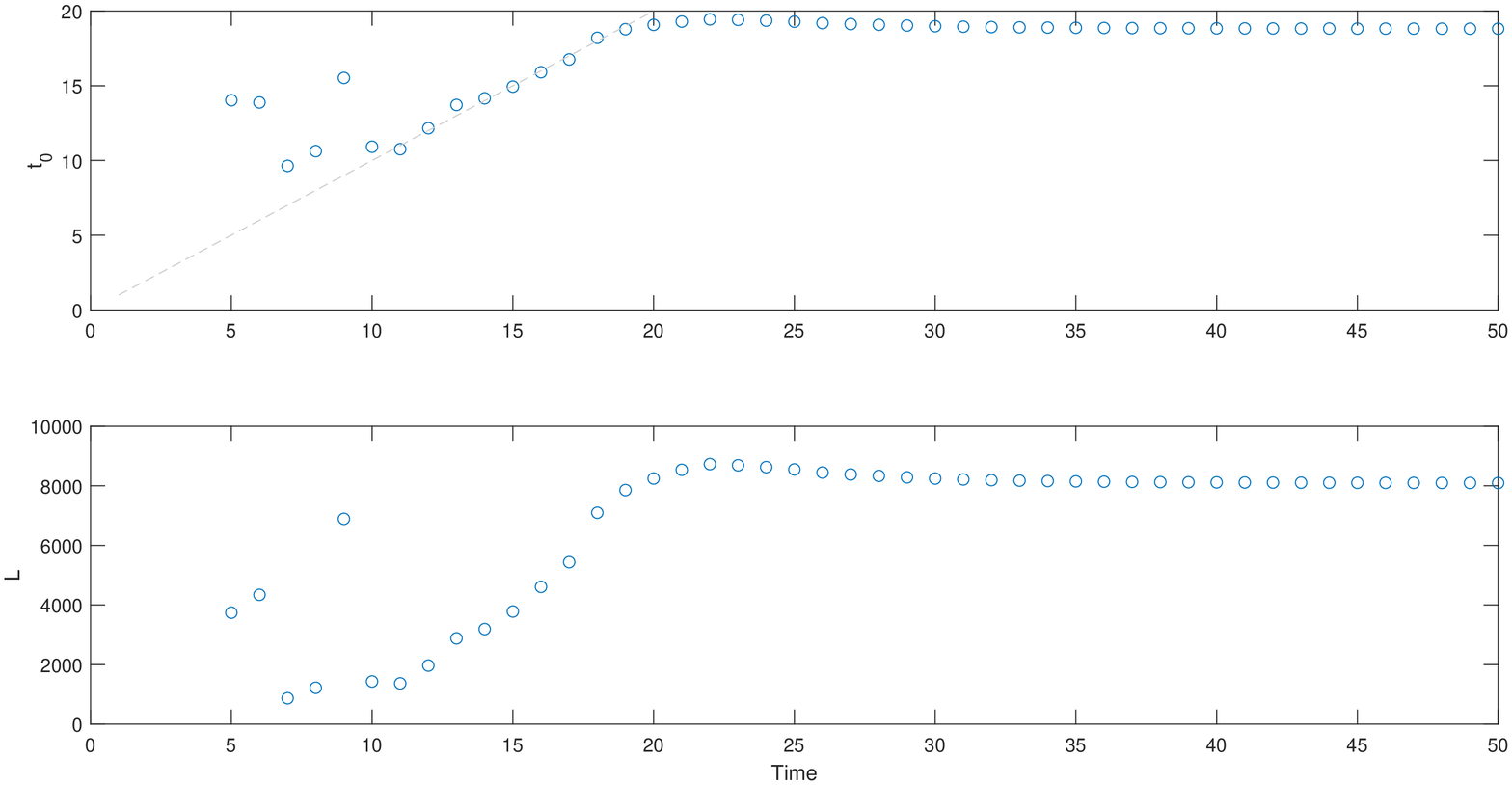}
    \caption{Example of instability of fit to the early part of data: plots of estimated inflection point and eventual case number as a function of when the forecast was made. Note how $t_0$ is attracted to the forecasting time (dashed grey line) until the inflection point around $t\approx 20$ arrives.}\label{fig:unidentifiable:2}
\end{figure}

It should not be surprising by now that the sigmoid family of curves show weak identifiability unless the whole range is shown. In a blog post by Martin Modr{\'a}k a sigmoid is indeed used to demonstrate weak identifiability for logistic curves \cite{Modrak2018identifying}. Seagren, Kim and Smets analysed the Andrews growth model of organisms in a bioreactor \cite{seagren2003identifiability}. This model can be solved explicitly and the sensitivity of the eventual substrate level to different parameters can be plotted. They found that  unique estimates of all parameters from a measurement can be obtained only within a narrow set of initial conditions: to actually fit the model several batch experiments have to be done. An agent based model of contagion shows no identifiability until in the middle course of the outbreak (see \autoref{fig:unidentifiable:1} and \autoref{fig:unidentifiable:2}).

A principled way of measuring the degree of identifiability is to calculate the Fisher information matrix, which describes the sensitivity of the model to changes in the data. If it is non-singular then the model is structurally identifiable \cite{rothenberg1971identification}. 
Daly et al. develop an effective Fisher information matrix that can be used to estimate a sensitivity matrix for a differential equation  \cite{daly2018inference}. They demonstrate it for a sigmoid growth model, finding that for early times the system is sensitive to changes in the growth rate $k$ but largely insensitive to changes in the asymptote $L$. As the inflection point is passed, the sensitivities flip: now data is effective at setting $L$ but not $k$. Thus the value of $k$ shows up well in early data but poorly in late data, and the reverse is true for $L$.


\begin{figure}
\begin{center}
\includegraphics[width=0.8\textwidth]{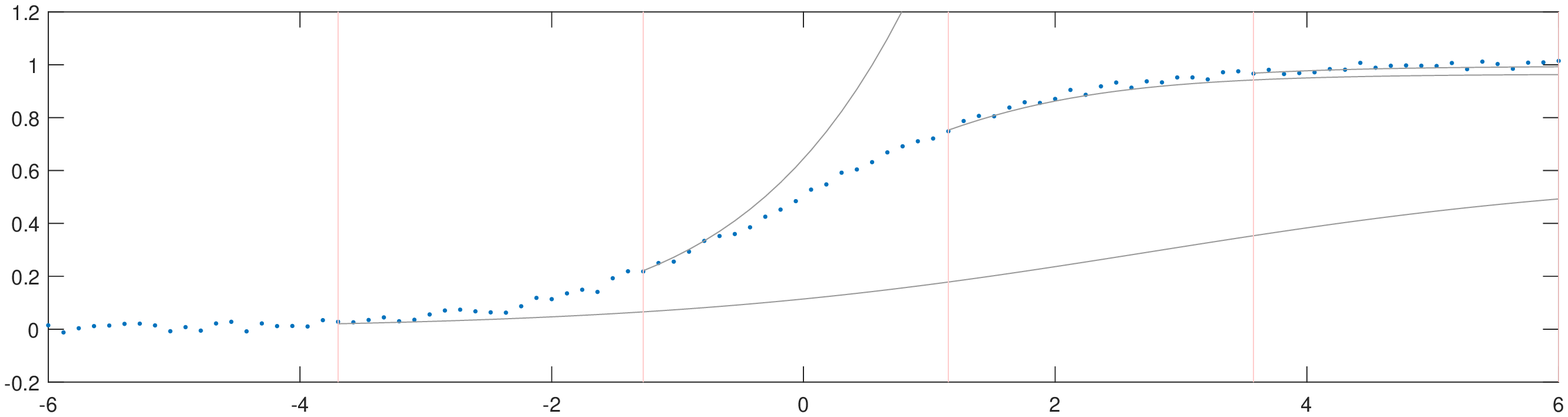}
\includegraphics[width=\textwidth]{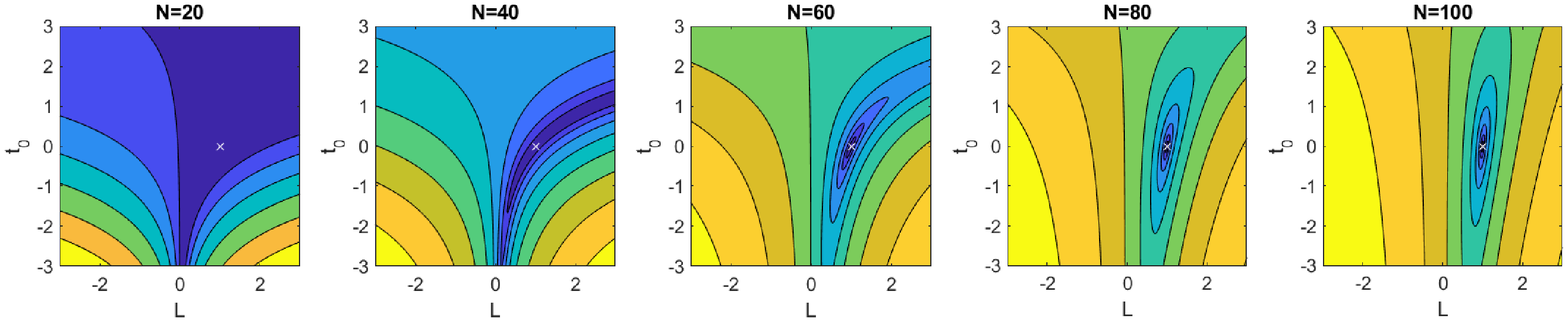}
\end{center}
\scaption{Magnitude of the mean square fitting error for different $L,t_0$ for a slightly noisy sigmoid at different time. A white 'x' marks the true values.
}
\label{fig:errorsurfaces}
\end{figure}
One can see this behaviour by plotting the mean square fit errors (the "error surface") as a function of different possible $L,t_0$ as more data arrives (figure \ref{fig:errorsurfaces}). Initially there is a vast range of equally good fits, showing the unidentifiability of the problem. Later there is a trench of equally good fits, still exhibiting unidentifiability. The problem becomes reminiscent of Rosenbrock's banana function, a common  test or demonstration for the difficulty of iterative methods to converge to the global minimum \cite{rosenbrock1960automatic}. As enough data arrives to make the system identifiable the global minimum becomes trivial to find.

\subsection{Identifiability and modelling}\label{modelling}

When designing a model and fitting it to data, it might be tempting to restrict to a family of curves like the logistic curves. Though those curves have weak identifiability, they only require three parameters to specify, and if those parameters can be estimated from early data, then their later behaviour can be identified.

But doing this opens up the issue of model uncertainty. Or, put more bluntly, if the initial data does not carry good $L$ and $(t_0,y_0=y(t_0))$ identifiability, then we cannot get this information simply by fitting a curve that does have identifiable $L$ and $(t_0,y_0)$. We would only get spurious certainty.

In the real world, early data might not carry strong information about the later shape of $y$ and $H$. For example, the classical technology adoption life cycle divides users into the categories of "innovators", "early adopters", "early majority", "late majority" and "laggards". This explicitly models the population as consisting of subgroups with different properties. And the inflection point is the transition from early to late majority; the asymptote is the sum of all the categories (including laggards).
Early data, though, comes entirely from innovators and early adaptors. If these populations are actually distinct in how they behave and how innovations spread among them, then the early data will tell us {\em nothing} about the inflection point and asymptote of the curve.

\begin{figure}
\begin{center}
\includegraphics[width=0.7\textwidth]{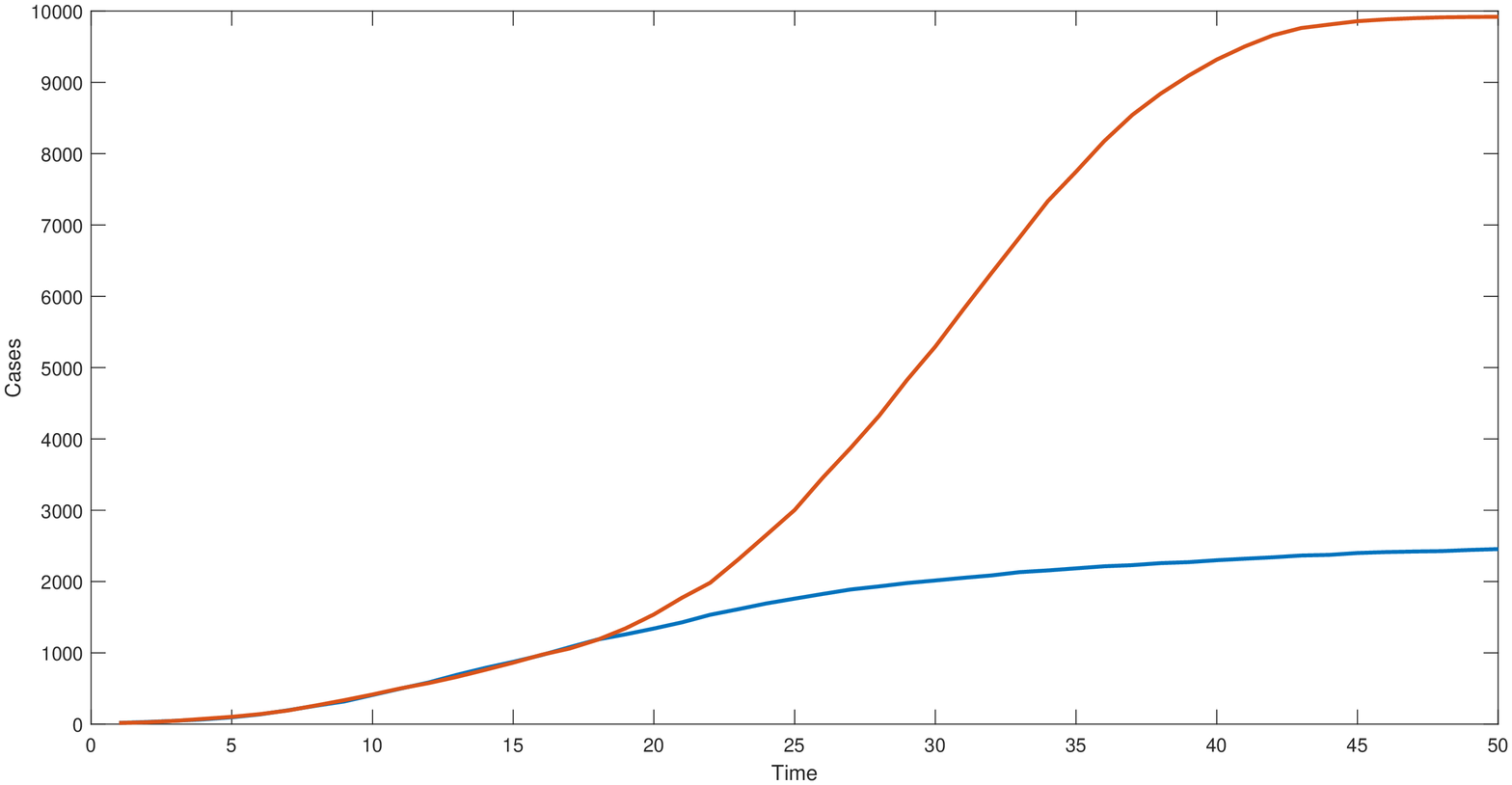}
\end{center}
\scaption{Example of unidentifiability in a growth model. Cumulative case numbers from two agent-based model of contagions. Each model has two populations: population $A$, $10\%$ of the population, and population $B$, $90\%$ of the population. Each population is defined by two parameters: their likelihood of immunity to the contagion (denoted $i_A$ and $i_B$) and incubation periods before they are infectious (denoted $\rho_A$ and $\rho_B$).
The blue curve has $i_A=30\%$, $i_B=80\%$, and $\rho_A = \rho_B = 1$ (population $A$ has $30\%$ immunity, population $B$ has $80\%$, and both have a one-day incubation period). The red curve has $i_A=10\%$, $i_B=0\%$, $\rho_A=1$, and $\rho_B=10$ (population $A$ has $10\%$ immunity and a one-day incubation period; population $B$ has $0\%$ immunity and a ten-day incubation period).
Until $t\approx 18$ there is no way of telling the curves apart and forecasts from the data would --- and should --- have been identical.
}
\label{fig:unidentifiable}
\end{figure}

An example of this is shown in an infection model in \autoref{fig:unidentifiable}. In this model, there are two populations (population $A$, $10\%$ of the population, and population $B$, $90\%$), and each population has a different level of immunity ($i_A$ and $i_B$) and a different incubation period ($\rho_A$ and $\rho_B$). The graph demonstrates that we could have two situations where these four parameters are quite different, but the infection curves nevertheless stay the same for $18$ days and then diverge very strongly.

\section{Remedies}

A trivial remedy to this form of forecasting error is ``don't do it'' --- do not rely on sigmoid fits of time series for forecasting.

While principled, it may also be overly pure for practical purposes. In science we can often afford to ignore weak and biased signals until they reach a statistically convincing level. However, no business or policy would be possible if we required $5\sigma^2$ evidence before taking any action. Decisions will be taken under uncertainty and always involve gambles: forecasts are information we can use to improve the odds.

The main remedy is to have as much domain knowledge as possible, giving good priors for fits and the ability to critique results.

\subsection{Knowing the model}

When fitting sigmoid models to data, it helps tremendously to have information about the dynamics that underlie the sigmoid.

\begin{figure}
    \centering
    \includegraphics[width=0.8\textwidth]{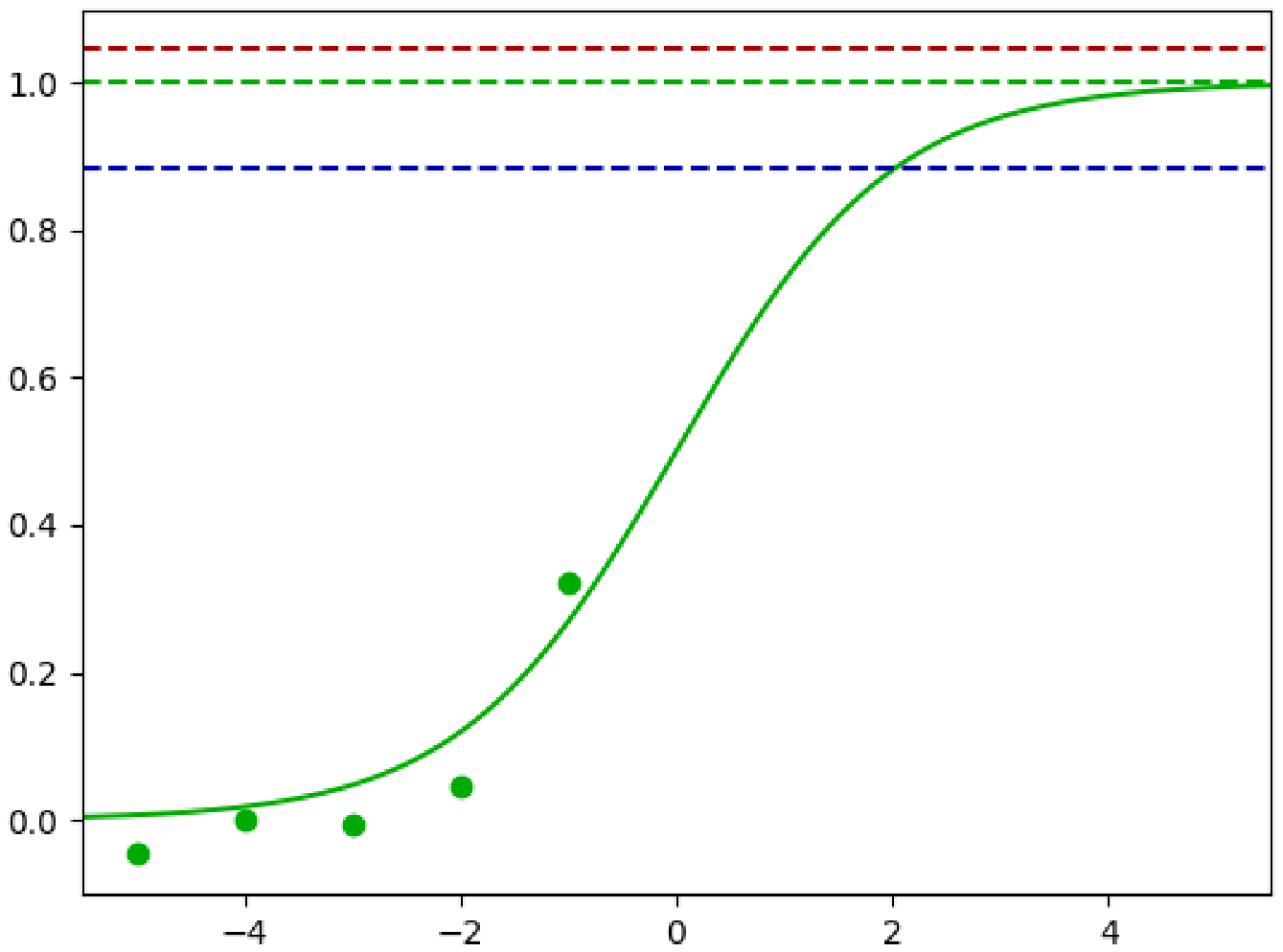}
    \caption{The true underlying data source is a logistic curve with $k=1$, $L=1$, and $t_0=0$ (the green curve). Five points are samples at $t\in\{-5,-4,-3,-2,-1\}$ with normally distributed noise of standard deviation $0.05$ (the five green points; their positions here are illustrative only). The value of $L$ is then estimated by a Bayesian update on these five samples. Repeating the process $500$ times gives an average estimate of $L=0.88$ when $k$ is unknown (the dotted blue line), and $L=1.04$ when $k=1$ is known (the dotted red line), which is closer to the true value $L=1$ (the dotted green line).
}
    \label{known_model}
\end{figure}

As a simple example, consider the situation presented in \Cref{known_model}. Here, the underlying curve is the simplest: a logistic curve with $k=1$, $L=1$, and $t_0=0$. We'll try and estimate the value of $L$, using an approach where $k$ is unknown, and a second where $k=1$ is known.

Five points are sampled with equidistant timing leading up to the inflexion point, with normally distributed noise (standard deviation of $0.05$). Then the parameters $k$, $L$, and $t_0$ are estimated in a Bayesian way; the prior is for $k$ to be uniform on $[0,4]$, $L$ uniform on $[0,2]$, and $t_0$ uniform on $[-4,4]$. These priors\footnote{
In practice, the values of $k$, $L$, and $t_0$ are used in discrete increments of size $1/50$.
} are then updated on the five samples.
That whole process, including the sampling, was repeated $500$ times, and the average posterior $L$ was computed to be $0.88$ (standard deviation of $0.14$). The whole process was also done for known $k=1$, which resulted in an average posterior $L$ of $1.04$ (standard deviation of $0.19$), which is better than twice as close to the true value $L=1$. 
However, note that this is the result of averaging 500 trials: for a single trial the uncertainty will be far larger.


Thus knowing more about the underlying dynamics --- in this case, the growth rate --- gives better estimates of the unknown parameters.

\subsection{Model known to be symmetric}

The asymptote $L$ is the value where $H(L)=0$; the inflection point has $y$ value at $y_0$ for which $yH(y)$ is maximised. The logistic curve has $y_0=L/2$, but it is clear that, for generic $H$, there is no reason to assume that $y_0$ and $L$ are related in such a way (see \Cref{diff:L:t0}).

But if $H(y)y$ is symmetric or roughly symmetric, then the $y_0$ value will be half of $L$, and the later part of the curve will the rotation of the early part around $(y_0,t_0)$ (the integral of an even function --- which all symmetric functions are --- is an odd function plus a constant term).

So, if $H(y)y$ is known to be symmetric, then once $H$ up to $y_0$ is known --- for example, once $t>t_0$ --- then $L$, and the later part of $y(t)$, can be well estimated.

\subsection{Known asymptote}

If the asymptote level is known, then one can estimate progress towards it from the near zero initial state. If this progress is sizeable, then it may be possible to estimate the inflection point.

\cite{tatrai2020covid} suggests the rule of thumb that when $y(t) > L/3$, then prediction becomes reliable. Given the large underestimates of actual values of $L$ in that work we can learn that even were the rule of thumb true, in many situations our initial estimates of $L$ may be badly wrong, resulting in posterior estimates that are equally spurious. Only when the level is given by independent evidence or principles (e.g. the problem is progress towards full saturation which will eventually happen, so $L$ is known) would such a rule be potentially trustworthy, and even then we may need knowledge of whether the sigmoid is symmetric or not in order to properly estimate.

For example, a disease cannot infect more people than the total population $P$, so we know that $L\leq P$. For a SIR model of disease, $L$ can be inferred from the $R_0$, the number of people each infected person infects. In ideal situations, this would give 'herd immunity' once $L_{herd}=P(1-1/R_0)$ people are infected:
That formula assumes a homogeneous population that mixes randomly with each other. One can get improved $L$ estimates by looking at similar diseases, estimating the network of spread, and so on. Unfortunately, in the early day of an epidemic when this information is most valuable, estimates of $R_0$ will also be highly uncertain (as well as models of transmission) if it is a new disease\footnote{The effective value of $R$ also depends on interventions such as social distancing that may depend on policy reactions to the model prediction. This problem of self-defeating forecasting is general, and separate from the modelling and mathematical problem this paper is dealing with. In the end, if the effective $R \ll R_0$ due to response to a forecast and the forecast eventually appears overly gloomy, this is a success! The point of forecasting should be to guide decisions to better outcomes than being right about the future.}.

\subsection{Unknown asymptote}

Without a bound on the asymptote the forecast is far more uncertain. 

It is useful is to recognize the tendency for early estimates of $L$ to gravitate towards twice the current level $y(t)$. In a sense it is a reasonable guess since we have a scale set by $y(t)$
and not knowing anything about $L$ we may assume it to be Pareto distributed (the Jeffreys prior for an unknown scale), giving $2y(t)$ as a best guess.\footnote{This may appear redolent of J. Richard Gott III's Copernican principle for estimating the remaining survival time of something as the current age \cite{gott1993implications}. However, that principle is based on the current data being uniformly distributed between $[0,L]$: for a sigmoid most data will be close to 0 or L, so in a sense being close to the half-way point of value is the least likely situation to find oneself in! It is still informative about how high $y(t)$ can rise. The principle {\em does} apply to estimating $t_0$ if a starting time is known, but should be applied with caution: see \cite{caves2000predicting,caves2008predicting} for critique and warnings.}
However, given the preceding treatment, we have regularly seen $L \gg 2y(t)$. Hence, heuristically, we should regard the estimated $L$ at  before the inflection point as a {\em lower bound} on the actual $L$. 

 This happens when the damping term grows linearly or faster as $y$ increases. If we have reason to believe this, the above heuristic makes sense. Conversely, if we suspect damping to grow fast as $y$ becomes non-zero we should regard the estimate as an upper bound. 
 
One approach suggested by Theodore Modis is to make several fits with different weights on the data points, then keeping the answer that gives the highest $L$ (most often the result of weighting the recent historical data points heavily, since these contain most information about late behaviour) \cite{modis2007strengths}.


\subsection{Estimating the inflection point}

Like $L$, the tendency for $t_0$ to be estimated close to the present can be used as a lower bound. This is unfortunately less useful since we only learn that we may not have passed it yet: a valuable caution, but uninformative.

In general $t_0$ cannot be estimated from the data ahead of time, because all we can detect for $t<t_0$ is that the rate of increase of $y(t)$ has decreased (which, if $y'(t)$ is differentiable, implies $y''(t)<0$). But this does not constrain when --- or even if --- $y'(t)$ will reach its maximum.
So without information about the dynamics of the model, our estimates of $t_0$ from initial data are likely to be incorrect.


\section{Conclusions}

The sigmoid issue is a member of a larger set of problems in forecasting where models cannot identify the best fit to data. It is striking because of its simplicity, and because it shows up again and again in the literature.

The problem also applies to more complex methods. Many of the more elaborate models in forecasting contain hidden or explicit sigmoid fits. In recent years neural networks have been suggested as forecasting tools. Unless there is a vast amount of data so the still-mysterious overparametrization powers of the networks can be unleashed, a well regularized network attempting to model a growing time-series is basically trying to pick up the same parameters as the sigmoid curve fit, and will have similar problems \footnote{An interesting case is "sloppy models", a universality class of nonlinear models with numerous parameters that are not identifiable but where the overall dynamics and fit to data is good. Their behaviour depends only on a few stiff combinations of parameters \cite{waterfall2006sloppy}. However, for forecasting applications there still has to be enough data to identify the stiff parameters.}.

Performing predictions and forecasts {\em well} does not mean one is never wrong. It also does not mean one states the most likely outcome(s) with appropriate uncertainty estimates, although this is a good start. The best thing to do is to give decision-relevant information that allows the receiver to update their model of their world to fit reality better and appropriately changes how they act. This is why the tendency for sigmoid models to place inflection points close in time (often a decision-relevant issue, as in considering responding to peak oil or epidemics) is pernicious and as a forecaster one should guard against it. The uncertainty about eventual asymptote level affects long-term planning (how much room is there to grow a company or economy? how bad will the pandemic be in total? where will Moore's law end up?) but in a softer way since it is rare for the exact magnitude to matter and it is expected to arrive further in the future. But sometimes, as in predicting herd immunity, the policy implications of a biased estimate can be harsh. 

Sigmoids are valuable tools, but as forecasting tools they have known biases and should hence be used carefully. Past performance is no indication of future results. 

\subsection{Acknowledgements}

We would like to thank commenters on LessWrong giving useful input, especially Aaro Salosensaari.
We also got valuable input from Cosma Shalizi.

This research was funded by the European  Research  Council (ERC) under the European Union’s Horizon 2020 research and  innovation  programme (grant  agreement  No.  669751), and by the Machine Intelligence Research Institute (MIRI).

\small
\bibliographystyle{plain}
\bibliography{sigmoids.bib}

\begin{thebibliography}{10}

\bibitem{Modrak2018identifying}
Identifying non-identifiability, 2018.
\newblock
  \url{https://www.martinmodrak.cz/2018/05/14/identifying-non-identifiability/}.

\bibitem{armstrong2001extrapolation}
J~Scott Armstrong.
\newblock Extrapolation for time-series and cross-sectional data.
\newblock In {\em Principles of forecasting}, pages 217--243. Springer, 2001.

\bibitem{bass1969new}
Frank~M Bass.
\newblock A new product growth for model consumer durables.
\newblock {\em Management science}, 15(5):215--227, 1969.

\bibitem{brandt2007testing}
Adam~R Brandt.
\newblock Testing hubbert.
\newblock {\em Energy Policy}, 35(5):3074--3088, 2007.

\bibitem{caves2000predicting}
Carlton~M Caves.
\newblock Predicting future duration from present age: A critical assessment.
\newblock {\em Contemporary Physics}, 41(3):143--153, 2000.

\bibitem{caves2008predicting}
Carlton~M Caves.
\newblock Predicting future duration from present age: Revisiting a critical
  assessment of gott's rule.
\newblock {\em arXiv preprint arXiv:0806.3538}, 2008.

\bibitem{crozier2020forecasting}
Constance Crozier.
\newblock Forecasting s-curves is hard, 2020.
\newblock
  \url{https://constancecrozier.com/2020/04/16/forecasting-s-curves-is-hard/}.

\bibitem{daly2018inference}
Aidan~C Daly, David Gavaghan, Jonathan Cooper, and Simon Tavener.
\newblock Inference-based assessment of parameter identifiability in nonlinear
  biological models.
\newblock {\em Journal of The Royal Society Interface}, 15(144):20180318, 2018.

\bibitem{kucharavy2011application}
Roland De~Guio Dmitry~Kucharavy.
\newblock Application of s-shaped curves.
\newblock {\em Procedia Engineering}, 9:559–572, 2011.

\bibitem{fotache2020predicting}
Chris Fotache.
\newblock Predicting the spread of covid-19 coronavirus in the us (live
  updates), 2020.
\newblock
  \url{https://medium.com/analytics-vidhya/predicting-the-spread-of-covid-19-coronavirus-in-us-daily-updates-4de238ad8c26}.

\bibitem{goldenberg2019s}
Matt Goldenberg.
\newblock S-curves for trend forecasting, 2019.
\newblock LessWrong 23rd Jan 2019
  \url{https://www.lesswrong.com/posts/oaqKjHbgsoqEXBMZ2/s-curves-for-trend-forecasting}.

\bibitem{gott1993implications}
J~Richard Gott~III.
\newblock Implications of the copernican principle for our future prospects.
\newblock {\em Nature}, 363(6427):315--319, 1993.

\bibitem{harvey1984time}
AC~Harvey.
\newblock Time series forecasting based on the logistic curve.
\newblock {\em Journal of the Operational Research Society}, 35(7):641--646,
  1984.

\bibitem{hsieh2004sars}
Ying-Hen Hsieh, Jen-Yu Lee, and Hsiao-Ling Chang.
\newblock Sars epidemiology modeling.
\newblock {\em Emerging infectious diseases}, 10(6):1165, 2004.

\bibitem{hubbert1956nuclear}
M~King Hubbert et~al.
\newblock Nuclear energy and the fossil fuel.
\newblock In {\em Drilling and production practice}. American Petroleum
  Institute, 1956.

\bibitem{lee2020estimation}
Se~Yoon Lee, Bowen Lei, and Bani Mallick.
\newblock Estimation of covid-19 spread curves integrating global data and
  borrowing information.
\newblock {\em PloS one}, 15(7):e0236860, 2020.

\bibitem{massiani2015choice}
J{\'e}r{\^o}me Massiani and Andreas Gohs.
\newblock The choice of bass model coefficients to forecast diffusion for
  innovative products: An empirical investigation for new automotive
  technologies.
\newblock {\em Research in Transportation Economics}, 50:17--28, 2015.

\bibitem{meade1998technological}
Nigel Meade and Towhidul Islam.
\newblock Technological forecasting—model selection, model stability, and
  combining models.
\newblock {\em Management science}, 44(8):1115--1130, 1998.

\bibitem{meade2001forecasting}
Nigel Meade and Towhidul Islam.
\newblock Forecasting the diffusion of innovations: Implications for
  time-series extrapolation.
\newblock In {\em Principles of forecasting}, pages 577--595. Springer, 2001.

\bibitem{modis2007strengths}
Theodore Modis.
\newblock Strengths and weaknesses of s-curves.
\newblock {\em Technological Forecasting \& Social Change}, 74(6):866--872,
  2007.

\bibitem{panik2014growth}
Michael~J Panik.
\newblock {\em Growth curve modeling: theory and applications}.
\newblock John Wiley \& Sons, 2014.

\bibitem{raeside1988use}
Robert Raeside.
\newblock The use of sigmoids in modelling and forecasting human populations.
\newblock {\em Journal of the Royal Statistical Society: Series A (Statistics
  in Society)}, 151(3):499--513, 1988.

\bibitem{rosenbrock1960automatic}
HoHo Rosenbrock.
\newblock An automatic method for finding the greatest or least value of a
  function.
\newblock {\em The Computer Journal}, 3(3):175--184, 1960.

\bibitem{rothenberg1971identification}
Thomas~J Rothenberg.
\newblock Identification in parametric models.
\newblock {\em Econometrica: Journal of the Econometric Society}, pages
  577--591, 1971.

\bibitem{san2017s}
JR~San~Crist{\'o}bal.
\newblock The s-curve envelope as a tool for monitoring and control of
  projects.
\newblock {\em Procedia computer science}, 121:756--761, 2017.

\bibitem{sandberg2014sigmoid}
Anders Sandberg.
\newblock A sigmoid dialogue, 2014.
\newblock \url{http://aleph.se/papers/A Sigmoid Dialogue.pdf}.

\bibitem{seagren2003identifiability}
EA~Seagren, H~Kim, and BF~Smets.
\newblock Identifiability and retrievability of unique parameters describing
  intrinsic andrews kinetics.
\newblock {\em Applied microbiology and biotechnology}, 61(4):314--322, 2003.

\bibitem{shen2020logistic}
Christopher~Y Shen.
\newblock Logistic growth modelling of covid-19 proliferation in china and its
  international implications.
\newblock {\em International Journal of Infectious Diseases}, 96:582--589,
  2020.

\bibitem{taleb2019probability}
Nassim~Nicholas Taleb, Duncan Needham, and Julius Weitzdorfer.
\newblock Probability, risk, and extremes.
\newblock {\em Extremes}, 31:46, 2019.

\bibitem{tatrai2020covid}
D{\'a}vid T{\'a}trai and Zolt{\'a}n V{\'a}rallyay.
\newblock Covid-19 epidemic outcome predictions based on logistic fitting and
  estimation of its reliability.
\newblock {\em arXiv preprint arXiv:2003.14160}, 2020.

\bibitem{wang2020prediction}
Peipei Wang, Xinqi Zheng, Jiayang Li, and Bangren Zhu.
\newblock Prediction of epidemic trends in covid-19 with logistic model and
  machine learning technics.
\newblock {\em Chaos, Solitons \& Fractals}, 139:110058, 2020.

\bibitem{waterfall2006sloppy}
Joshua~J Waterfall, Fergal~P Casey, Ryan~N Gutenkunst, Kevin~S Brown,
  Christopher~R Myers, Piet~W Brouwer, Veit Elser, and James~P Sethna.
\newblock Sloppy-model universality class and the vandermonde matrix.
\newblock {\em Physical review letters}, 97(15):150601, 2006.

\bibitem{young1993technological}
Peg Young.
\newblock Technological growth curves: a competition of forecasting models.
\newblock {\em Technological forecasting and social change}, 44(4):375--389,
  1993.

\end{thebibliography}

\appendix

\section{Sigmoid definition}\label{sigmoid:def:sec}

We can formally define a sigmoid as:

\begin{definition}[Sigmoid]\label{sigmoid:def}
A sigmoid with growth rate $k>0$, inflection point $t_0\in\mathbb{R}$, and asymptote $L>0$, is a function $y$ such that:
\begin{enumerate}
    \item $y$ is defined on $(\infty,T)$, where $T$ is either a real number or $T=\infty$.
    \item $y(t)>0$ and $y(t)<L$ for all $t$ in its domain.
    \item $\lim\limits_{t\to -\infty} y(t)=0$ and $\lim\limits_{t\to T} y(t) = L$
    \item $y(t)$ is strictly increasing.
    \item $y(t)$ is differentiable.
    \item $t=t_0$ is the unique local maximum of $y'(t)$.
    \item $\lim\limits_{t\to -\infty} y'(t)=0$ and $\lim\limits_{t\to T} y'(t)=0$ (making $t=t_0$ into the global maximum for $y'(t)$).
    \item $\lim\limits_{t\to -\infty} y'(t)/y(t)=k$.
\end{enumerate}
\end{definition}

Note that this definition includes sigmoids that reach $L$ in finite time, as well as those that asymptote to $L$ over infinite time.

\section{Differential equation results}\label{diff:appendix}

Recall that $H(y)$ is decreasing with $H(0)=1$, $H(L)=0$, and $H(y)>0$ for $y<L$. The function $H(y)y$ has a single global maximum on $(0,L)$, at $y=y_0$.

Thus $\frac{1}{k H(y)y}$ is defined and strictly positive on $(0,L)$. It is monotone increasing for $y>y_0$ and monotone decreasing for $y>y_0$. Thus it is integrable on any closed subinterval of $(0,L)$

The equation $y'(t) = k \cdot H(y(t)) \cdot y(t)$ is separable, giving $dy/(kH(y)y) = dt$. So solutions are given by the integral equation:
\begin{eqnarray}
\int _{y(0)}^{y(t)}\frac{1}{k H(y^*)y^*} dy^* = \int_0^t dt^* = t,
\end{eqnarray}
using $y^*$ and $t^*$ as dummy integration variables.

Fixing any specific value for $y(0)$ in $(0,L)$, that integral equation has a solution. Moreover, since $\frac{1}{k H(y^*)y^*}$ is strictly positive, that solution is strictly increasing in $y(t)$. Thus, once $y(0)$ is fixed, $y(t)$ is strictly defined as a function of $t$ --- and we can also see $t$ as a function of $y$.

Fix $t' < y(0)$. On the range $(0,y(t'))$, $H(y) > \delta$ for some $\delta>0$. We also have that $H(y)<2$. So
\begin{eqnarray}
\int _{y(0)}^{y(t')}(1/ \delta k)\frac{1}{y^*} dy^* < \left( t' = \int _{y(0)}^{y(t')}\frac{1}{k H(y^*)y^*} dy^* \right) < \int _{y(0)}^{y(t')}(1/2 k) \frac{1}{y^*} dy^*.
\end{eqnarray}
Since $\int 1/y^* dy^*$ diverges at $y=0$, this demonstrates that $t'\to -\infty$ as $y(t')\to 0$. Now define $T$ as the limit of $t$ as $y(t)\to L$; since $1/\frac{1}{k H(y)y}$ diverges as $y(t) \to L$, this might be a finite real $T$, or $T=\infty$.

Thus $y(t)$ has domain $(-\infty, T)$. It is clear that such a $y(t)$ is strictly increasing and differentiable with derivative $y'(t)=k \cdot H(y(t)) \cdot y(t)$. By assumption on $H(t)$, $y'$ has a unique local and global maximum $y_0$ on $[0,L]$; define $t_0$ by $y(t_0)=y_0$. And clearly $y'(t) \to 0$ as $t$ tends to $-\infty$ or $T$. Since $H(0)=1$, $\lim_{t\to-\infty}y'(t)/y(t)=\lim_{t\to-\infty} kH(y(t)) = kH(0)=k$.

Thus $y(t)$ is a sigmoid in the sense of \Cref{sigmoid:def}

Conversely, let $y(t)$ be a sigmoid in the sense of \Cref{sigmoid:def}; then it must be differentiable. Then define $H(y)$ as:
\begin{eqnarray}
H(y(t)) = \frac{y'(t)}{ky(t)}.
\end{eqnarray}
The sigmoid properties of $y(t)$ then insure that $H(0)=1$, $H(L)=0$, $H(y)>0$ for $y<L$, and $H(y)y$ has a single global maximum on $(0,L)$.

\end{document}